\documentstyle[pre,aps]{revtex}
\input{psfig.sty}
\begin{document}
\draft
\twocolumn[\hsize\textwidth\columnwidth\hsize\csname@twocolumnfalse\endcsname
\title{Interaction of Vortices in Complex Vector Field and Stability of a
``Vortex Molecule''}
\author{Igor S. Aranson$^1$ and Len M.~Pismen$^2$}
\address{$^{1}$Argonne National Laboratory, 9700
South Cass Avenue, Argonne, IL 60439\\
$^2$Department of Chemical Engineering, Technion, 32000 Haifa, Israel}
\date{\today}
\maketitle
\begin{abstract}
We consider interaction of vortices  in the vector
complex Ginzburg--Landau equation (CVGLE).
In the limit of small field coupling,
it is found analytically that the interaction
between  well-separated defects in two different fields is
long-range, in contrast to interaction between
defects in the same field which falls off exponentially.
In a certain region of
parameters of CVGLE, we find stable rotating bound states of
two defects -- a ``vortex molecule".
\end{abstract}
\vskip1pc]

The complex Ginzburg--Landau equation (CGLE)
is  a paradigm model for qualitative description of weakly
nonlinear oscillatory media (see for review \cite{Cross}). This equation is
a generic form
which is obtained as the amplitude equation in the vicinity of a
Hopf bifurcation. There is a vast literature on vortex  solutions of
CGLE and on interactions and instabilities of vortices
\cite{Cross,oup}.

Recently, the attention has been brought to the vector extension of CGLE.
The problem of nonlinear dynamics of a 2D complex vector field arises most
naturally in the context of nonlinear optics, where the order parameter is
the electric field envelope in the plane normal to the direction of
propagation \cite{Gil,P94,Hae}; the fields $A_\pm$ can be identified with
the two circularly polarized waves of opposite sense. Alternatively,
coupled complex fields can be interpreted as amplitudes of interacting
nonlinear waves \cite{Bridges}. A distinguished feature of the vector
Ginzburg--Landau equation (VGLE) is a possibility of transition between two
``phases'', which can be characterized by either ``mixing'' or
``separation'' of two ``superfluids''. Defects (vortices) can exist in both
``superfluids'', and transitions between alternative core structures are
possible \cite{oup,P94}; in this sense, VGLE could be viewed as a toy model
of $^3$He dressed down to two dimensions.

A particularly intriguing possibility, suggested  in Ref.~\cite{P94}, is
the formation of a
bound pair of defects in the two fields, i.e. a vortex ``molecule'' with
dipole structure. We shall show in this Letter that such a ``molecule''
cannot in fact exist in the model with real coefficients, but is readily
formed in the vector model with complex coefficients (CVGLE). The latter
form appears, in particular, as an amplitude
equation near the lasing transition \cite{Gil,max95}. Recent simulations of
CVGLE \cite{max98} have shown spiral wave patterns with an exceptionally
rich structure where both separated (but closely packed) defects in the two
fields and ``vector'' defects with a common core (called argument and
director vortices in Ref.~\cite{P94}) could be seen. There has been so far,
however, no theoretical studies of isolated defects in CVGLE and their
interactions.

In this Letter  we study
interaction of defects in CVGLE in the limit of small coupling between
two complex fields. We have found that the  interaction
between a well-separated pair of defects in two different fields is
always long-range (power-like), in contrast to the  interaction
between defects in the same field which falls off exponentially as in a
single CGLE \cite{akw}. In a certain region of
parameters of CVGLE we found stable rotating bound states of
two defects -- a "vortex molecule". Analytical results are in excellent
agreement with simulations.

Under appropriate scaling of
the physical variables, the vector equation for two symmetric interacting
complex fields $A_\pm$ acquires the universal form
\begin{eqnarray}
\partial_t A_\pm&=&A_\pm -(1+ic)\left(|A_\pm|^2+g |A_\mp|^2 \right) A_\pm
\nonumber \\ &+&(1+ib) \nabla^2 A_\pm,
\label{cgle0}
\end{eqnarray}
where real parameters $b$ and $c$ are, respectively, the  ratio of
dispersion  to diffusion and the ratio of conservative to dissipative
nonlinearity, and  the complex parameter $g=g_r + i g_i$
characterizes the magnitude of the coupling.

A scalar defect of Eq.~(\ref{cgle0}) with unit topological charge is a
one-armed  spiral in  $A_+$ (or $A_-$) field, while a vector
defect is formed by two scalar defects with a common core. Simple energy
considerations, applicable in the case of VGLE with real coefficients,
point out that scalar defects in different fields tend to separate or
unite, respectively, when $g$ is positive or negative. For the CVGLE, no
energy integral can be defined, but the limit $g \to 0$ is, of course,
distinguished, since the interaction ceases, and defects in both fields,
described by the usual Hagan's \cite{hagan} rotating spiral solution,
are mutually independent.

{\it Defects in two different fields.}
At $|g| \ll 1$ the interaction can be treated perturbatively.
For $g=0$
the scalar defect of Eq.~(\ref{cgle0}) in either field centered at the
origin is
\begin{eqnarray}
A (r,\theta) &=& F(r) \exp i [ -\omega t + \theta + \psi(r) ],
\label{spir}
\end{eqnarray}
where $(r, \theta)$ are polar coordinates, $\omega = c +(b-c)k^2_0$,
are  the rotation frequency, and $k_0$ is an asymptotic wavenumber
emitted by the spiral. The
real functions $F$ and $\psi$
have the following asymptotic behavior  for $r \to \infty$:
$F \to  \sqrt{1 -k_0^2 }$
and $\chi \to   k_0 $,
where $\chi=\psi^\prime( r)$ is a local wavenumber.
The  wavenumber $k_0$
is determined uniquely
for given $b,c$ \cite{hagan}.

A form more convenient for the analysis,  obtained after
setting $ A_\pm =u_\pm \sqrt{(1+\omega b)/(1+bc)}e^{-i\omega t}$
and rescaling
$
\nabla \to \nabla \sqrt{(1+b\omega)/(1+ b^2)}$, $
\partial_t \to (1+b\omega) \partial_t$,
is
\begin{eqnarray}
(1-ib) \partial_t u_\pm&=&(1+i\Omega)u_\pm \nonumber \\
&-&(1+iq)\left(|u_\pm|^2+ g
|u_\mp|^2 \right) u_\pm + \nabla^2 u_\pm,
\label{cgle}
\end{eqnarray}
where $q=(c-b)/(1+bc)$, $\Omega =(\omega-c)/(1+\omega c)$.

Due to the interaction, the positions of defects
${\bf r}_\pm (t)$ become  slow functions of time, so that
the instantaneous  drift velocity ${\bf \dot{r}_\pm}\equiv
{\bf v}_\pm = O(|g|)$.
Rewriting Eq.~(\ref{cgle})
in the comoving frame, we obtain in the first order
\begin{eqnarray}
&&(1-ib) {\bf v}_\pm\cdot \nabla u_\pm +(1+i\Omega)u_\pm \nonumber  \\
&-&(1+iq)\left(|u_\pm|^2+  g |u_\mp|^2 \right) u_\pm + \nabla^2
u_\pm=0,
\label{cgw}  \end{eqnarray}
The imaginary part of the advective
term can be absorbed by transforming $ u_\pm \to u_\pm \exp
\left(\frac{1}{2}i b{\bf v_\pm \cdot x}\right)$, which accounts
for the Doppler shift in the emitted wave\cite{comm}.

We concentrate upon one defect, say, that marked by the index $+$ (which we
will further omit), take its position as the origin and view interaction
with its counterpart as a perturbation. Expanding in $g $, we write
\begin{eqnarray}
u_\pm &=&
 \left(F(|{\bf r}-{\bf r}_\pm|)+  w_\pm \right)
 \exp i [\theta_\pm + \psi(|{\bf r} -{\bf r}_\pm |) ],
\label{spir_l1}
\end{eqnarray}
where ${\bf r}_\pm $ are positions of the zeroes of the respective fields
and $\theta_\pm $ are  polar angles about
these points; $w_\pm $ is an $ O(|g|)$ correction.
Substituting the ansatz (\ref{spir_l1}) into Eq.~(\ref{cgw}), we obtain the
first-order equation
${\cal H} + \Psi=0$,
containing the linear operator ({\it cf.} \cite{akw})
\begin{eqnarray*}
{\cal H}= \! -(1+iq) F^2 ( w +w^*) + \Delta w
+ \frac{2 i }{r^2} \partial_\theta w
 +2 i \chi F\partial_r \left( \frac{ w}{F} \right)
\end{eqnarray*}
and the inhomogeneity
\begin{eqnarray*}
\Psi =-(1+iq) g F F^2 (\bar r)
+  ( F^\prime+ iF \chi) {\bf v \cdot n }- i \frac{F}{ r} {\bf v \times  n }.
\end{eqnarray*}
Here  $\bar r =|{\bf r -R}|$, ${\bf R=r_- -r_+}$ is the separation between the
defects, ${\bf n}=(\cos \theta, \sin \theta)$ is the unit vector along {\bf
R}, and $ \Delta = \nabla^2 - F^{-1}\nabla^2 F$.

The operator ${\cal H}$ has two Goldstone modes corresponding to the
translational symmetry in the plane. Therefore the inhomogeneous equation
${\cal H} + \Psi=0$ has bounded solution  if $\Psi$
is orthogonal to the adjoint zero modes $ w^+$
of the homogeneous problem. Thus, the drift velocity can be
derived from the solvability condition (see Ref.\cite{oup})
\begin{equation}
\mbox{Re} \int w^+({\bf r}) \Psi({\bf r} ) d^2{\bf r} =0,
\label{sc}  \end{equation}
The operator ${\cal H}$ is not self-adjoint
for any $q \ne 0$. As a result, the adjoint zero modes cannot be expressed
through the translational modes $\nabla u_0$, and have to be computed
directly by solving the equation ${\cal H}^+ ( w^+ ,w^{+*}) =0$.
The
adjoint operator ${\cal H}^+$ is
\begin{eqnarray}
{\cal H}^+ = -F^2 ( w^+ +w^{+*} - iq (w^+-w^{+*})) +
 \Delta w^+ \nonumber \\
- \frac{2 i }{r^2} \partial_\theta w^+
 + \frac{2 i }{rF} \partial_r \left(r\chi F w^+ \right) .
     \label{LA}     \end{eqnarray}

We take note that the Goldstone modes of Eq.~(\ref{cgle}) are first
harmonics $ w = e^{-i (\theta + \psi)}\nabla u_0$. 
In view of the
orthogonality of the eigenfunctions of the operator and its adjoint with
distinct eigenvalues, the zero modes of the adjoint operator must contain
the first harmonics as well. To solve the equation ${\cal H}^+ =0$, we first
separate the real and imaginary parts of $w^+=t^++i s^+$
and then take the latter's
first harmonics $(t^+, s^+)= \left (T^+(r ), S^+(r )\right)
e^{i \theta} + c.c$. Thus, we
derive from Eq.~(\ref{LA})
\begin{eqnarray}
\Delta_r T^+& + &
\frac{2 i S^+ }{r^2}
-\frac{2}{rF} (r \chi  F S^+)^\prime
- 2  F^2 (T^+ + q S^+) =0,
 \nonumber \\
\Delta_r S^+& - &
  \frac{2 i T^+ }{r^2}  +
 \frac{2}{r F} (r \chi F T^+)^\prime =0 ,
\label{lin10}
\end{eqnarray}
where $\Delta_r= \partial^2_r +r^{-1} \partial_r - r^{-2}- F^{-1}\nabla^2 F$.
A typical structure of the adjoint mode in shown in Fig.~1.
One can find that
the adjoint mode decays exponentially for $r \to \infty$
(see also Ref.~\cite{abk}).

\begin{figure}
\centerline{\hspace{.0cm} \psfig{figure=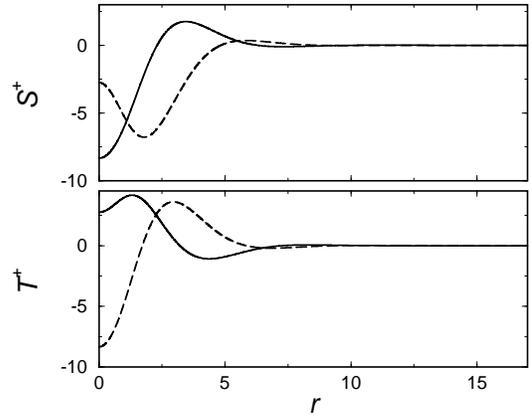,height=2.5in}}
\caption{
The real  and imaginary parts (solid/dashed  lines) of $T^+$ and $S^+$
components of the adjoint mode for $q=1.3$.
 }
\end{figure}

The solvability condition (\ref{sc}) can be rewritten in a compact form by
defining the complex velocity $v =v_x -i v_y$, so that ${\bf v \cdot n}
=\mbox{Re}(ve^{i\theta}),\; {\bf v \times  n }=\mbox{Im}(ve^{i\theta})$.
Assuming that {\bf R} is directed along the $x$ axis, $v_x$ and  $v_y$
coincide with the radial and tangential velocity components, respectively,
$v_r$ and  $v_\tau$. Expressing also the first harmonics of the
cross-coupling term through the angular integral $ \bar F(r ) = \pi^{-1}
\int_{0}^{\pi} \cos \theta
(F(\bar r ))^2 d \theta$, we derive from Eq.~(\ref{sc})
the equation for the complex velocity
\begin{equation}
v= {\cal I}^{-1} \gamma_r \int_0^\infty \left( T^+ + \frac{\gamma_i}{
\gamma_r}  S^+\right)F \bar F( r) r dr .
\label{vel3}
\end{equation}
where $\gamma \equiv\gamma_r +i\gamma_i = (1 +iq)g $, and the
friction factor is
\begin{eqnarray*}
{\cal I}= \frac{1}{2} \int_0^\infty \left[ F^\prime (r) T^+
+\left( \chi F+\frac{ iF }{r} \right) S^+ \right] r dr .
\end{eqnarray*}

This expression is the principal analytical result of our work. One
can see immediately that, with a generic complex $T^+(r )$ and $S^+(r
)$, the velocity can be modified simply by rotating the argument of the
complex
interaction parameter $g$. In this way, one can ensure that the
radial component $v_r=\mbox{Re}(v)$
vanishes at some finite $R$ at least in a
certain interval of $\arg(g)$. If this equilibrium
position is stable, it signals the formation of a bound state, which we
call a {\it vortex molecule\/}. Since at the equilibrium distance  the
tangential component of
the velocity given by $v_\tau={\rm Im} (v)$ is generally non-zero, this
bound state must rotate with a certain angular velocity.

The  bound states cannot form in the real VGLE.
In this case the operator ${\cal H}$ is
self-adjoint,  and the adjoint mode is just the translation
mode,  $(T^+, S^+) = (F^\prime, -iF/r)$. From  Eq.~(\ref{vel3}) one derives
then $v_r \sim g \int F F^\prime \bar F  r dr $ and $v_\tau=0$.
Since $F( r)$ is a monotonic function,  $v_r$ does not change sign.
One finds that the vector defect is unstable for $g>0$ and stable
otherwise. The eigenvalue $\lambda$ of the mode
responsible for splitting the vector defect as function of $g$ is shown in Fig.~4
(inset).

For $R\gg 1 $ and $R\ll 1 $ Eq.~(\ref{vel3})
can be calculated analytically. Using the asymptotic expansions
$F^2(\bar r)  \approx 1-k^2 + k_0 /(q \bar r) +... $  valid for $R \gg1 $
we obtain
\begin{equation}
\bar F \approx
\frac{ k_0 }{ \pi q } \int_{0}^{\pi} \frac{\cos \theta d \theta }
{\sqrt{r^2 + R^2 -2 r R \cos \theta } }
 \approx \frac{ k_0 r }{2  q R ^2 }.
\label{mu1}
\end{equation}
For $R\ll 1$
one uses  $F(\bar r)= F(r)+F(r)^\prime(R/r- R \cos \theta)$,
leading to
$\bar F =- R  FF^\prime(r ) $.
The solvability condition (\ref{vel3}) yields then
$v= - R \gamma_r   \alpha_1$ for $R\ll 1$,
$v =   R^{-2 } \gamma_r \alpha_2$ for $R\gg 1$ with
the constants $\alpha_{1,2}$  given by
\begin{eqnarray}
 \alpha_1&=&
{\cal I}^{-1} \int_0^\infty F^2 F^\prime(r ) \left
( T^+ + \frac{\gamma_i}{\gamma_r} S^+\right)
r dr,
\nonumber \\
 \alpha_2&=&
\frac{ k_0}{2q}\,{\cal I}^{-1}\int_0^\infty F(r ) \left
( T^+ + \frac{\gamma_i}
{\gamma_r} 
S^+\right) r^2 dr .
\label{sol3}
\end{eqnarray}

\begin{figure}
\centerline{\hspace{.0cm} \psfig{figure=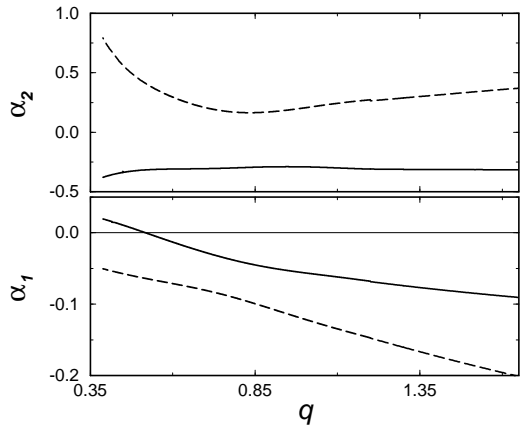,height=2.5in}}
\caption{The real (solid line) and imaginary (dashed  line) parts of
$ \alpha_1$ and $ \alpha_2$ as functions of $g$ (for $g$ real).
 }
\end{figure}

Provided $F(r),\chi $ and the adjoint mode $T^+,S^+$ are known, the
dependence of velocity on distance   $R$  can be
found explicitly.
We used a matching-shooting algorithm to find a stationary spiral solution
and a corresponding adjoint eigenmode. Then, we evaluated numerically the
integrals in Eq.~(\ref{vel3}) and calculated both
the constants $\alpha_{1,2}$ (Fig.~2) and
radial 
and tangential velocities  
as functions of the distance $R$ (Fig.~3).
We find that for $g_i=0$,  $g_r>0$  and $q<q_c\approx 0.52$
the defects repel one another ($v_r<0$) at small distances,
so that a vector defect is unstable, as in the case of real VGLE.
Surprisingly, for larger $q>q_c$ the real part of
$\alpha_1$ changes sign, and
the defects bind at $R=0$ forming a stable vector defect.
Correspondingly, for $g_r<0$ the vector defect is stable
at $q<q_c$ as for real VGLE, but becomes unstable at $q>q_c$.

\begin{figure}
\centerline{\hspace{.0cm} \psfig{figure=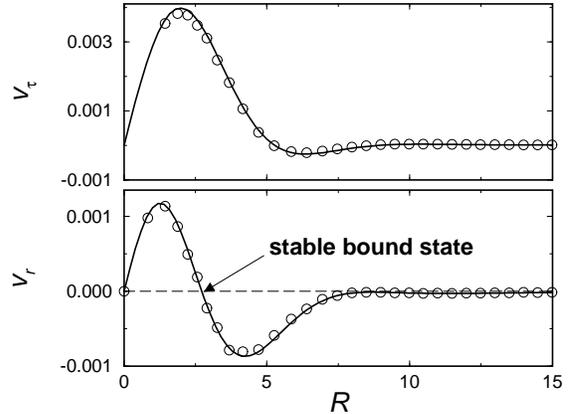,height=2.5in}}
\caption{
The radial and tangential velocities $v_r,\, v_\tau$ as functions of the
separation distance
$R$ for $q=1.3$ and $g =-0.01$. The solid line has been computed using
Eq.~(\protect\ref{vel3}),
circles present results of simulations of Eq.~(\ref{cgle}).
 }
\end{figure}
\begin{figure}
\centerline{\hspace{.0cm} \psfig{figure=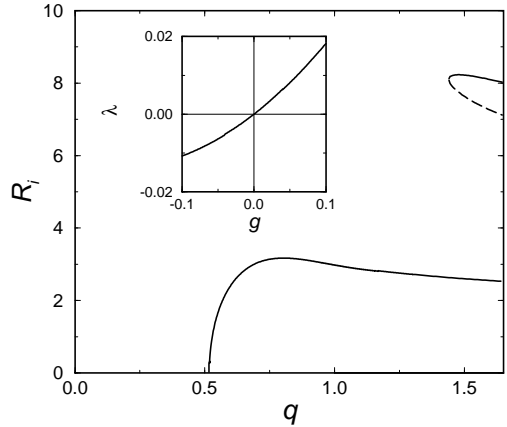,height=2.5in}}
\caption{
Equilibrium separation  $R$  {\it vs.}  $q$ obtained from
Eq.~(\protect \ref{vel3}). For $g_r<0$ and $g_i=0$ solid lines indicates stable
radii and dashed line unstable ones.
Inset: The eigenvalue $\lambda$ {\it vs.}  $g$ for real VGLE.
 }
\end{figure}
The bifurcation at $q=q_c$ is subcritical for $g_r>0$ but supercritical for
$g_r<0$.
In the latter case, it generates a  stable solution with finite $R$.
Under these conditions, the dependence
of the radial velocity $v_r$ on $R$ has a zero at some equilibrium
distance, as shown in Figs.~3, 4.
For $g_r>0$ this bound state is unstable.
As $q$ further increases, a new pair of stable/unstable
bound states
emerges as a result of a saddle-node bifurcation. This is shown in Fig.~4.
The number of equilibrium radii is
finite since the asymptotic behavior of the radial velocity is always
monotonic.
Certainly, only the bound state
with the smallest radius is important, since at large $R$ the ``binding
strength"
decreases.

The analytical results were compared with numerical
simulations of Eq.~(\ref{cgle}). We used the  Crank-Nicholson method
in the domain of $250\times250$ units with $500\times 500$ grid points
and
with non-reflecting boundary conditions (see for details \cite{chate}).
The analytical and numerical results appears to be in excellent agreement,
as seen in Fig.~3.

{\it Defect pair  in the same field.}
The interaction of oppositely-charged defects in the same field
can be reduced to interaction of a single defect with a plane boundary
\cite{akw}. The problem of interaction in the scalar CGLE is determined
by the growing exponential solution of the stationary linearized problem
(see for details Ref.~\cite{akw}). As a result,
the velocity due to interaction with the boundary is given by
$v \sim e^{-pR}$, where $R$ is the distance to the boundary and $p$ is
the root with the positive real part
of the corresponding characteristic equation obtained
at $r\gg 1$.  As shown in Ref. \cite{akw},
$p$ is  real for $q\le 0.85$ and complex for $q>0.85$. Thus,
interaction of the defect with a plane boundary (or two defects) is
oscillatory, and a variety of (unstable) bound states is possible.

\begin{figure}
\centerline{\hspace{.0cm} \psfig{figure=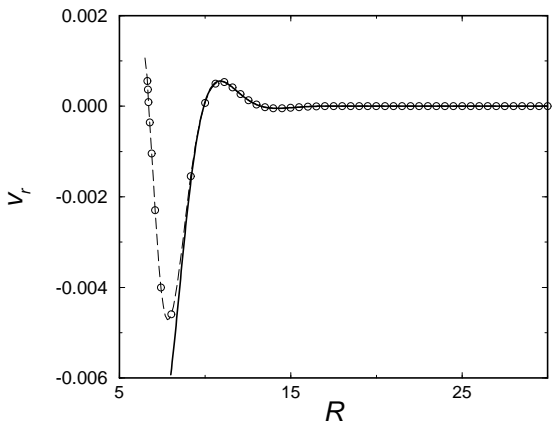,height=2.5in}}
\caption{
The radial velocity $v_r$ {\it vs.}  distance to the boundary $R$
for $q=1.5$ and $g=0.04$.
The dashed line with circles shows the results of simulations
with Eq.~(\protect\ref{cgle0}),
the solid line is a fit by $v \sim  e^{-p R} + c.c $, where
$p=0.746+i 0.94$ is the root of the
characteristic equation of the stationary problem
 \protect \cite{akw}.
 }
\end{figure}

The interaction problem in the framework of CVGLE is similar
to that of the scalar CGLE. Looking at
stationary perturbations of CVGLE $w_{\pm}$, we find that
generic  solutions grow exponentially away from the defect core
$w_\pm \sim A_0 e^{p r}$, where $p$ is the corresponding
root of the
characteristic equation and $A_0$ is the eigenvector. Thus, we expect
exponentially decaying interaction of the defects in CVGLE,
$v \sim e^{-pR}$. For $g \to 0$ the
corresponding exponents should be close to those of the scalar CGLE.
This exponential interaction is not surprising, since the waves emitted by
the defects collide and form shocks which screen  the cores of the defects.

This exponential interaction is verified by numerical simulation
of CVGLE. As one sees in Fig.~5, the dependence of the radial velocity
(i.e. the component along the line connecting the cores)
{\it vs.} $R$ is very well approximated by an
exponential dependence with the exponent derived from the solution of
the stationary problem.

We have shown that the
interaction between two well-separated  defects
in different complex field is always long-range. In a certain parametric
domain,
these defects may form stable rotating bound states.
In contrast, the interaction
between defects in the same field falls off exponentially.
Our analytical results are limited to a small coupling  constant $g$.
Additional core instabilities may be encountered as $g$ grows
\cite{max98}.

We are grateful to Maxi San Miguel and Lorenz Kramer
for illuminating discussions.
ISA   acknowledges support from the U.S. DOE under grants
W-31-109-ENG-38 and
NSF, STCS \#DMR91-20000. LMP acknowledges hospitality of the
Max-Planck-Institut f\"ur Physik Komplexer Systeme,
Dresden and Fritz-Haber-Institut, Berlin, and support by the Fund for
Promotion of Research at  the Technion and the Minerva
Center for Nonlinear Physics of Complex Systems.

\end{document}